\documentclass[twocolumn]{aastex631}

\newcommand{\reviewhighlight}[1]{#1}

\usepackage{amsmath}
\usepackage{savesym}
\savesymbol{tablenum}
\usepackage{siunitx}
\restoresymbol{SIX}{tablenum}
\usepackage{multirow}

\hypersetup{linkcolor=red,citecolor=magenta,urlcolor=blue}

\received{TBD}
\revised{\today}
\accepted{TBD}

\submitjournal{ApJ}

\shorttitle{POPULATION III STARS}
\shortauthors{LARKIN et al.}

\graphicspath{{./}{figures/}}

\begin{document}

\title{Characterization of Population III Stars with Stellar Atmosphere and Evolutionary Modeling and Predictions of their Observability with the James Webb Space Telescope}

\correspondingauthor{Mikaela M. Larkin}
\email{mmlarkin16@gmail.com}

\author{Mikaela M. Larkin}
\affiliation{Center for Astrophysics and Space Sciences, University of California San Diego, La Jolla, CA 92093, USA}

\author[0000-0003-0398-639X]{Roman Gerasimov}
\affiliation{Center for Astrophysics and Space Sciences, University of California San Diego, La Jolla, CA 92093, USA}

\author[0000-0002-6523-9536]{Adam J.\ Burgasser}
\affiliation{Center for Astrophysics and Space Sciences, University of California San Diego, La Jolla, CA 92093, USA}

\begin{abstract}

Population III stars were the first stars to form after the Big Bang, and are believed to have made the earliest contribution to the metal content of the universe beyond the products of the Big Bang Nucleosynthesis. These stars are theorized to have had extremely short lifespans, and therefore would only be observable at high redshifts ($z \geq 3-17$) and faint apparent magnitudes ($m_{AB} \gtrsim 40$). The direct detection of Population III stars therefore remains elusive. However, the recently launched James Webb Space Telescope (JWST) may be capable of detecting stars in the relevant magnitude range in the event of favorable gravitational lensing. Theoretical models are required to interpret these future observations. In this study, new evolutionary models and non-equilibrium model atmospheres were used to characterize the observable properties of zero-age main sequence Population III stars. The calculated models cover a wide range of possible Population III stellar masses, from the minimum mass predicted by star formation studies to the maximum mass capable of maintaining hydrostatic equilibrium. Synthetic photometry and theoretical color-magnitude diagrams were calculated for the bands of the Near-Infrared Camera (NIRCam) on JWST. The final results are compared to the scales of known lensing events and JWST magnitude limits. The purpose of this study is to calculate the observable parameters of Population III stars in the most optimal JWST bands in order to provide a theoretical foundation for anticipated future observations of this stellar population.



\end{abstract}


\section{Introduction} \label{sec:intro}

Population III stars are an elusive addition to the two traditional stellar populations identified in \citet{baade_populations}. This yet unobserved population accommodates the earliest and nearly metal-free stars that formed shortly after the Big Bang. Nearly seven decades ago, \citet{first_popIII_1} recognized that early stars were likely more massive, luminous and short-lived compared to their present-day Population I and II counterparts. The need for a non-standard formation mechanism to produce such stars and the relatively high metallicity measured in the most metal-poor Population II stars known at the time ($[\mathrm{Fe}/\mathrm{H}]\approx -3$, \citealt{popIII_search_1}) led \citet{first_popIII_2}\footnote{Note that the argument in \citet{first_popIII_2} is based on the assumption that dark matter is composed of low-mass stars and brown dwarfs which has been largely ruled out \citep{no_BD_DM_1,no_BD_DM_2,no_BD_DM_3}.} and \citet{Wagner} to independently identify the earliest stars as a distinct stellar population. In addition to the earliest chemical enrichment of the primordial gas, Population III stars may have contributed to the cosmic microwave background \citep{CMB_distortion}, the cosmic infrared background \citep{CIB_0,CIB_1,CIB_2,CIB_3}, the gravitational wave background \citep{GW_1,GW_2,GW_3}, reionization and reheating of the universe \citep{reionization_1,reionization_2,reionization_3,reionization_4,reionization_5,reheating} and likely had a noticeable feedback effect on the formation of the first galaxies \citep{galaxy_formation_1,galaxy_formation_2}.

Recent searches for extremely metal-poor stars have revealed a sharp cut-off in the metallicity distribution around $[\mathrm{Fe}/\mathrm{H}]\approx -4$, with fewer than $50$ known objects below this cut-off \citep{JINAbase}. These include the most metal-poor star known ($[\mathrm{Fe}/\mathrm{H}]= -6.2$, \citealt{most_metal_poor_star}) as well as at least one object with an unmeasured iron abundance and the estimated upper limit  $[\mathrm{Fe}/\mathrm{H}] \lesssim -7$ \citep{even_more_metal_poor_star}. However, the comparatively high abundances of other chemical elements (e.g. $[\mathrm{C}/\mathrm{H}]\gg -3$ for both stars) challenge the status of these sources 
as true Population III representatives, unless the observed abundances are acquired from the interstellar medium through selective accretion, as proposed by \citet{pollution}.

The lack of identifiable metal-free stars in surveys is consistent with the long-standing theoretical expectation of a top-heavy initial mass function (IMF) that precludes primordial stars from having sufficiently long lifespans to survive until the present day. This expectation is motivated by the lack of metals in the early universe, which leaves $\mathrm{H}_2$ and $\mathrm{HD}$ as the only available cooling agents in molecular clouds, thereby suppressing fragmentation and producing unusually heavy stars \citep{popIII_IMF,population_transitions}. Early numerical simulations (e.g. \citealt{heavy_popIII_1,heavy_popIII_2,heavy_popIII_4}) suggested that Population III stars predominantly formed with masses in excess of $100\ M_\odot$ and collapsed into black holes, with the exception of a subset of stars with masses between $140\ M_\odot$ and $260\ M_\odot$ that produced pair-instability supernovae \citep{heavy_popIII_3,Heger}. Later studies with a more detailed treatment of radiative feedback, interactions between stars, turbulence, etc., have challenged this picture by allowing formation of $\gtrsim 10\ M_\odot$ Population III stars (e.g. \citealt{light_popIII_1,light_popIII_2,light_popIII_3,light_popIII_4}) or even $< 1\ M_\odot$ stars (e.g. \citealt{ultralight_popIII_1,ultralight_popIII_2,ultralight_popIII_3,light_popIII_5}) that may exist in the present-day Milky Way, likely disguised by metal-enriched mass transfer from heavier stars \citep{AGB_popIII}. The existence of lower-mass primordial stars enables additional mechanisms of chemical enrichment, potentially explaining the observed inconsistency of the abundance patterns in metal-poor stars with predictions of pair-instability supernova yields \citep{inconsistent_pair_instability}. A distinct hypothetical population of supermassive primordial stars with masses in excess of $10^3-10^6\ M_\odot$ has also been proposed as seeds for the supermassive black holes in high-redshift quasars \citep{SMS_1,SMS_3,SMS_2,SMS_4}.

Since the majority of Population III stars are expected to have masses between a few tens and a few hundred solar masses, the correspondingly short lifespans ($\lesssim 20\ \mathrm{Myr}$, \citealt{Windhorst}) necessitate direct observation of such objects at high redshifts. The first stars begin to form once the primordial molecular clouds, concentrated around growing dark matter over-densities, cool down sufficiently to become unstable against gravitational collapse. The star formation rate as a function of redshift can be traced in simulations of cosmological hydrodynamics \citep{popIII_high_z,popIII_high_z_1,popIII_high_z_2,Jaacks}. The first Population III stars are expected to form around $z\approx30$ ($\sim0.1\ \mathrm{Gyr}$ after the Big Bang) and the maximum formation rate density ($\sim 10^{-4}-10^{-3}\ M_\odot\ \mathrm{yr}^{-1}\ \mathrm{Mpc}^{-3}$) is attained around $17\lesssim z \lesssim 10$ ($\sim0.2-0.5\ \mathrm{Gyr}$ after the Big Bang). This result is generally consistent with early reionization optical depth measurements from the Wilkinson Microwave Anisotropy Probe (WMAP) \citep{OmegaM,WMAP_1,WMAP_2}, although later WMAP \citep{WMAP_4} and Planck \citep{WMAP_5} measurements have cast doubt on the usability of this parameter as a probe of primordial star formation \citep{WMAP_6}. Population III stars give way to Population II stars once the metal mass fraction ($Z$) of the interstellar medium reaches a critical value, $Z_\mathrm{cr}\gtrsim 10^{-8}-10^{-6}$ ($Z_\mathrm{cr}\gtrsim 10^{-6}Z_\odot-10^{-4}Z_\odot$\footnote{Solar metallicity taken as $Z_\odot = 0.01$ to the nearest order of magnitude.}; \citealt{crit_Z_0,crit_Z_1,crit_Z_2,crit_Z_3,crit_Z_4}) that allows for efficient cooling and fragmentation of collapsing molecular clouds. The lowest redshift at which Population III stars may be observed remains uncertain as isolated metal-free pockets may last for extended periods of time, producing new Population III stars at later epochs. Population III star formation is expected to continue until at least $z=6$ ($\sim1\ \mathrm{Gyr}$ after the Big Bang, \citealt{popIII_low_z_1,popIII_low_z_2}) and possibly much later under special circumstances \citep{popIII_low_z_3}. Searching for metal-free stars at redshifts as low as $z=3$ ($\sim2\ \mathrm{Gyr}$ after the Big Bang) is particularly important given observations of the Lynx arc -- a star forming region at $z=3.4$ with a Population III-consistent ionization source \citep{Fosbury}; and LLS1249 -- a dense gas cloud at $z=3.5$ with a Population III remnant metallicity \citep{popIII_remnants}.

A renewed interest in direct detection of individual Population III stars has developed in anticipation of observations with the next generation of ground- and space-based facilities, in particular the recently launched James Webb Space Telescope (JWST). Using the Main Sequence properties of Population III stars from the evolutionary calculations in \citet{Schaerer_models} and new metal-free model atmospheres calculated with the \texttt{TLUSTY} code \citep{TLUSTY}, \citet{Rydberg_JWST} estimated the observable properties of isolated primordial stars at a range of redshifts and stellar masses. Without gravitational lensing, Population III stars appear far too faint to be detected even at the lowest redshift ($z=2$) and in extremely long exposures ($100\ \mathrm{hr}$). \citet{Rydberg_JWST} also considered the case of a favorable lensed observation through the galaxy cluster MACS J0717.5$+$3745 -- one of the largest gravitational lenses known \citep{largest_lens}. Even in the lensed case, a realistic detection was found to require either an extremely heavy Population III candidate ($\geq\ 300\ M_\odot$) or a primordial star formation rate, far in excess of theoretical expectation.

However, very high magnifications ($\mu$) may be attained for brief periods of time in the event of gravitational lensing during a caustic transit. A caustic is the locus of points in the source plane where the determinant of the magnification matrix vanishes, i.e. where a true point source would experience infinite magnification \citep{intro_to_lensing}. A compact light source such as an individual star crossing a caustic may experience extreme magnification up to $\mu\sim 10^7$ from a lensing cluster with a continuous distribution of mass under most favorable conditions \citep{max_magnification}. In practice, microlenses within the galaxy cluster will distort the lens caustics, reducing the maximum magnification to $\mu\sim 10^4$ \citep{diego_2018,diego_2019}. 

Since larger magnifications require more favorable and increasingly less likely configurations, the true expected magnification in any given survey will strongly depend on the redshift of interest, the number of observable targets, and the scope of the survey itself. For example, \citet{popIII_magnification} calculate $\mu\gtrsim 700$ as a realistic magnification estimate for detecting Population III stars in a $100\ \mathrm{deg}^2$ ultra-deep survey. Extreme lensing events with $\mu\gg 10^3$ have allowed for recent discoveries of the most distant individual stars known at $z=1.5$ (``\textit{Icarus}'', $\mu>2000$, \citealt{most_distant_star}), $z=2.7$ (currently unnamed, $\mu\gtrsim 10^4$, \citealt{another_most_distant_star}) and $z=6.2$ (``\textit{Earendel}'', $\mu>4000$, \citealt{even_more_distant_star,Earendel_2}). It has been suggested that the last source may in fact be a Population III star \citep{possible_popIII_detection}.

Adopting more optimistic magnifications of $\mu\sim 10^4 - 10^5$, \citet{Windhorst} used new metal-free evolutionary tracks calculated with the \texttt{MESA} (\texttt{M}odules for \texttt{E}xperiments in \texttt{S}tellar \texttt{A}strophysics) code \citep{MESA,MESA_2,MESA_3,MESA_4,MESA_5} and assumed blackbody emergent spectra to characterize Population III stars in the context of future observations with JWST. The study estimated that a decade-long observational program monitoring up to $30$ candidate lensing clusters will be required for a reliable detection.

In this study, we contribute to the ongoing effort of predicting future observations of Population III stars using stellar modelling. In particular, we focus on primordial stars with initial masses between $1\ M_\odot$ and $\sim10^3\ M_\odot$ on the Zero Age Main Sequence (ZAMS). ZAMS properties of Population III stars provide a lower bound on observability since later evolutionary stages are intrinsically more luminous and subjected to less interstellar absorption due to redder colors \citep{Schaerer_models,Windhorst}. Additionally, the lower surface gravities of post-ZAMS stars may lead to super-Eddington luminosities in high-mass candidates, requiring detailed modelling of mechanical motion in the atmosphere that falls beyond the scope of this study (however, see \citealt{winds_1,Yoon}).

We present theoretical color-magnitude diagrams of ZAMS Population III stars in JWST NIRCam filters for a broad range of redshifts based on new metal-free evolutionary models and model atmospheres. The observable properties of primordial stars are analyzed for their dependence on individual opacity sources and non-equilibrium distribution of the radiation field throughout the atmosphere. In particular, we demonstrate that even in the absence of non-grey opacity sources, Population III atmospheres cannot be approximated as blackbodies and always require detailed modelling. The calculated metal-free physical parameters on ZAMS are compared to their metal-poor counterparts and matched to simple, physically-motivated analytic relationships. The models are also evaluated in the context of the Eddington limit, that is of particular importance at high initial masses.

In this paper, Section~\ref{sec:methods} describes our modelling toolkit and presents the new atmosphere and evolutionary models calculated in this study. The key physical properties of Population III stars inferred from the models, such as the dependence of stellar evolution on the dominant energy production mechanism and the Eddington limit, are discussed in this section as well. Section~\ref{sec:results} details our color-magnitude calculations at high redshift and presents the predicted color-magnitude and mass-magnitude relationships for Population III stars in the context of future JWST observations with gravitational lensing. The effect of cosmological parameters on our predictions is also estimated in Section~\ref{sec:results}. The key findings and important shortcomings of this investigation are summarized in Section~\ref{sec:conclusion}. Population III model parameters are tabulated in Appendix~\ref{sec:appendix}.

\section{Modeling} \label{sec:methods}

\subsection{Overview of the methodology}

In this study, predictions of colors and magnitudes of Population III stars are drawn from synthetic photometry of metal-free stellar models at ZAMS. Each model is parameterized exclusively by the initial stellar mass and must account for all relevant physical processes governing the evolution of the star from its pre-Main Sequence (PMS) phase until steady-state hydrogen fusion. Synthetic photometry is obtained from the evolved emergent spectrum of the star, which is, in turn, calculated by solving the radiative transfer equation at every wavelength throughout the outer layers of the model comprising the stellar atmosphere.

Stellar atmospheres are particularly challenging to model due to the presence of neutral and partially ionized species, resulting in complex, wavelength-dependent opacity from significant non-grey contributions of bound-free and bound-bound sources. At high effective temperatures and extremely low metallicities considered in this work, the effect of non-grey atmospheric opacity on the structure of the stellar interior is expected to be insignificant, avoiding the need for detailed opacity calculations in the evolutionary models. However, the atmospheric opacity must be re-introduced into the model when calculating the final emergent spectrum of the star. We therefore calculate all models in multiple stages. First, evolutionary modelling was carried out from PMS to ZAMS with grey atmospheric opacity. The evolved stellar radii and luminosities were used to derive simple analytic relationships between stellar mass and the ZAMS surface parameters (effective temperature, $T_\mathrm{eff}$, and surface gravity, $\log_{10}(g)$). Finally, the analytic relationships were evaluated at a broad range of stellar masses (from $1\ M_\odot$ to $1000\ M_\odot$) and the resulting surface parameters were used as inputs to dedicated model atmosphere calculations with the full opacity treatment and spectral synthesis.

\subsection{Evolutionary Modeling}

We calculated all evolutionary models with the \texttt{MESA} code \citep{MESA,MESA_2,MESA_3,MESA_4,MESA_5}, version 15140. Evolutionary calculations in \texttt{MESA} are carried out in adaptive time steps until the chosen termination condition is met. By default, structure equations at each step are solved using the grey atmosphere optical depth-temperature relationship as the surface boundary condition. As demonstrated in our previous work \citep{roman_omegacen}, this approximation is satisfactory at effective temperatures ($T_\mathrm{eff}$) above $5000\ \mathrm{K}$ -- a condition met by all models considered in this study.

In all evolutionary models, we chose to use $Y=0.25$ as the approximation for the primordial helium mass fraction based on the Planck measurement in \citet{H0}, taken to the nearest percent to match the default precision in \texttt{MESA}.

At initial masses below $90\ M_\odot$, models are initialized as PMS with a central temperature of $5\times 10^5\ \mathrm{K}$ (following \citealt{MIST,roman_omegacen}), uniform chemical composition, and a density profile that satisfies the structure equations and results in the desired stellar mass.
At significantly larger initial masses, a convergent PMS solution may not exist. Instead, the $90\ M_\odot$ PMS is used as the starting point and the mass is slowly increased to the required value using the mass relaxation routine provided by \texttt{MESA}. As an example, at the initial mass of $1000\ M_\odot$, the relaxation process lasts $\approx 1300\ \mathrm{yr}$ and results in an object with the central temperature of $72.7\times 10^6\ \mathrm{K}$.

We extract the ZAMS luminosity and stellar radius from all evolutionary models once their nuclear power output begins to exceed $90\%$ of the total luminosity. The sensitivity of the derived ZAMS parameters on the primordial helium mass fraction ($Y$) as well as the adopted PMS settings was estimated by computing multiple models at the representative initial masses of $10\ M_\odot$ and $1000\ M_\odot$ for a range of $Y$ values from $0.24$ to $0.26$; a range of initial central temperatures from $3\times 10^5\ \mathrm{K}$ to $7\times 10^5\ \mathrm{K}$; and a range of maximum (pre-relaxation) PMS masses from $50\ M_\odot$ to $100\ M_\odot$ (stored in the \texttt{max\_mass\_to\_create} variable of the \texttt{build\_pre\_ms\_model()} subroutine). We found the effect of PMS settings to not exceed $0.002\ \mathrm{dex}$ in both luminosity and radius at ZAMS. The effect of the chosen $Y$ value was slightly larger, reaching $0.02\ \mathrm{dex}$ at lower masses. However, neither of the aforementioned uncertainties exceeds the average accuracy of the calculated analytic mass-radius and mass-luminosity relationships (to be introduced below) that were estimated as $0.02\ \mathrm{dex}$ and $0.03\ \mathrm{dex}$ respectively. Therefore, our results are insensitive to the input parameters within the aforementioned ranges. 

We calculated $428$ evolutionary models with initial stellar masses ranging from $1\ M_\odot$ to $1000\ M_\odot$ and used the ZAMS radii and luminosities to derive analytical mass-radius and mass-luminosity relationships for Population III stars. The input settings (inlist files) for all evolutionary models are available \dataset[online]{https://zenodo.org/record/7145568\#.Yzzax33MLAU}\footnote{\href{https://zenodo.org/record/7145568\#.Yzzax33MLAU}{https://zenodo.org/record/7145568\#.Yzzax33MLAU}}. Radii of main sequence stars are generally well-described by power law relations assuming that the dominant energy production and transport mechanisms do not vary significantly  \citep{mass_radius_1,mass_radius_2,mass_radius_3}. For stars with solar metallicity, the power law index changes noticeably around $\sim 1\ M_\odot$ due to the dissipation of the outer convective zone \citep[Ch. 22.1]{ms_powerlaw_indices} and the transition of the main hydrogen fusion mechanism from the proton-proton chain to the carbon-nitrogen-oxygen (CNO) cycle \citep[Ch. 5.2]{Maurizio_book}. The latter effect is particularly important for zero-metallicity stars, as it is expected that enough carbon will be produced at sufficiently high masses to display a similar transition, thereby offsetting the power law break into the range of masses considered in this study. We therefore model the mass-radius relationship of ZAMS Population III stars as a broken power law with the break point mass ($M_\mathrm{bp}^{(R)}$) treated as a free parameter. The adopted relationship is

\begin{equation} \label{eq:1}
    R\propto\begin{cases}
        M^{\alpha},& \text{if } M\leq M_\mathrm{bp}^{(R)}\\
        M^{\beta},& \text{if } M>M_\mathrm{bp}^{(R)}\\
    \end{cases}
\end{equation}

\noindent where $M$ is the initial stellar mass, $R$ is the corresponding stellar radius, and $\alpha$ and $\beta$ are the proton-proton and CNO power indices respectively. The mass-luminosity ($M$--$L$) relationship is slightly more complicated due to the dependence on the dominant pressure support in the interior. High-mass ($M\gtrsim 10\ M_\odot$) stars behave approximately as Eddington standard models ($n=3$ polytrope; \citealt{Edd_standard,ZAMS_massive}) with the mass-luminosity power index gradually changing from $\geq3$ at $M\lesssim100\ M_\odot$ to the asymptotic linear relationship ($L\propto M$) in the limit of $M\rightarrow \infty$. The transition occurs as the equation of state in the interior changes from the ideal gas law to radiation pressure dominance. To accommodate this additional complexity, we allow the index of the power law to change linearly with $\log_{10}(M)$ below some break point mass ($M_\mathrm{bp}^{(L)}$), which is treated as a free parameter, as before. The relationship is

\begin{equation} \label{eq:2}
    L\propto\begin{cases}
        M^{\delta\log_{10}\left(M/M_\odot\right)+A},& \text{if } M\leq M_\mathrm{bp}^{(L)}\\
        M^{\gamma},& \text{if } M>M_\mathrm{bp}^{(L)}\\
    \end{cases}
\end{equation}

Here, $L$ is the luminosity, $\gamma$ is the high-mass power law index, $\delta$ is the rate of change of the index in the low-mass regime, and $A$ is a constant, fixed by the requirement of the relationship to remain smooth around the break point,

\begin{equation} \label{eq:A}
    A=\gamma-2 \delta\log_{10}\left(M_\mathrm{bp}^{(L)}/M_\odot\right)
\end{equation}

The best-fit values of all six free parameters, as well as best-fit normalization factors, are given in Table~\ref{tab:best_fit}. Note that the standard ZAMS mass-radius relationship for Population I ($\mathrm{PI}$) stars has $\alpha_\mathrm{PI}\approx 0.8$ and $\beta_\mathrm{PI}\approx 0.57$ (\citealt{popper}, \citealt[Ch. 22.1]{ms_powerlaw_indices}). While the high-mass index matches the Population III value in Table~\ref{tab:best_fit}, the low-mass index is discrepant by a factor of $\sim 4$. This discrepancy arises because the energy production mechanism transition in Population I stars ($\approx 1.3\ M_\odot$; \citealt[Ch. 5.2]{Maurizio_book}) very nearly coincides with the onset of convection in the envelope ($\sim 1\ M_\odot$), while Population III stars maintain radiative envelopes on either side of $M_\mathrm{bp}^{(R)}$. To compare Population III and Population I mass-luminosity relationships, we consider the average power law index in the $1\leq M/M_\odot \leq 10$ and $1\leq M/M_\odot \leq 40$ ranges. The standard values for Population I are $3.88$ and $3.35$ respectively (\citealt{popper}, \citealt[Ch. 22.1]{ms_powerlaw_indices}), while the corresponding Population III values using Eq.~\ref{eq:2} and Table~\ref{tab:best_fit} are $3.87$ and $3.50$, suggesting that the Population III relationship is nearly identical to its Population I counterpart at low masses and marginally steeper at higher masses.

\begin{figure}
    \centering
    \includegraphics[width=1\columnwidth]{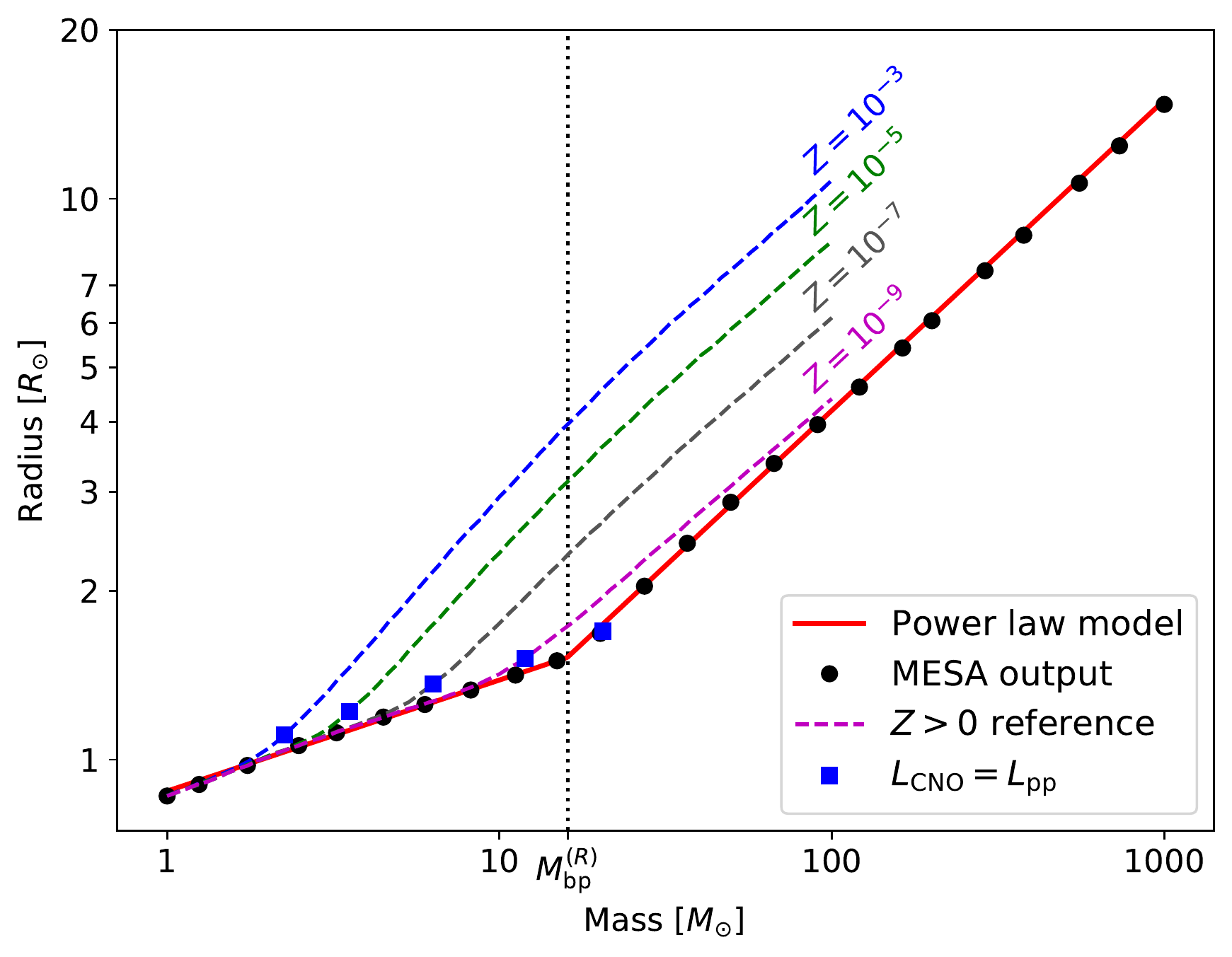}
    \caption{Mass-radius relationship from the metal-free evolutionary models calculated in this study alongside the best-fit power law approximation given in Eq.~\ref{eq:1} and using the best-fit parameters in Table~\ref{tab:best_fit}. Equivalent relationships for non-zero metallicities were calculated as well and are shown with dashed lines for comparison. Each dashed curve is parameterized by the total metal mass fraction, $Z$. The vertical line indicates the break in the best-fit power law relationship that originates from the onset of the CNO cycle in the core. The point where the energy production rates of the proton-proton chain ($L_\mathrm{pp}$) and the CNO cycle ($L_\mathrm{CNO}$) match is indicated with blue squares for every mass-radius relationship shown.}
    \label{fig:massvsrad}
\end{figure}

The mass-radius relationship is plotted in Fig.~\ref{fig:massvsrad} alongside the direct output from selected \texttt{MESA} models. For comparison, we also calculated multiple grids of \texttt{MESA} models at non-zero metallicities, whose mass-radius relationships are overplotted in the figure as well. The expected broken power law behavior is observed at all considered metallicities; however, the break point occurs at progressively decreasing stellar masses with increasing metallicity due to the larger initial carbon abundance. The point of equality in the energy production rates of the proton-proton chain and the CNO cycle is also indicated in the figure for every mass-radius relationship shown. Note that while the equality point is strongly correlated with the power law break point, the latter occurs at a lower stellar mass than the former as even a subdominant contribution from the CNO cycle is sufficient to influence the pressure structure within the star. In particular, we calculated the zero-metallicity equality point as $\approx 20.5\ M_\odot$ -- over $4\ M_\odot$ higher than $M_\mathrm{bp}^{(R)}$.

\begin{deluxetable}{lrlll}
\tablenum{1}
\tablecaption{Analytic fit parameters\label{tab:best_fit}}
\tablewidth{\columnwidth}
\tablehead{
\colhead{Parameter} & \colhead{Value} & \colhead{Error} & \colhead{} 
}
\startdata
   $\alpha$ & $0.1982$ & $\pm0.0019$&\\
   $\beta$ & $0.5527$ & $\pm0.0008$&\\
   $M_\mathrm{bp}^{(R)}$ & $16.03$ & $\pm0.11$  & $M_\odot$ \\
   $R$ at $1\ M_\odot$ & $0.8792$ & $\pm0.0034$  & $R_\odot$ \\
   \hline
   $\delta$ & $-0.6893$ & $\pm0.0030$&\\
   $\gamma$ & $1.3137$ & $\pm0.0080$&\\
   $M_\mathrm{bp}^{(L)}$ & $153.0$ & $\pm2.8$& $M_\odot$ \\
   $L$ at $1\ M_\odot$ & $1.850$ & $\pm0.023$  & $L_\odot$ \\
\enddata
\end{deluxetable}



The effective temperature ($T_\mathrm{eff}$) and surface gravity ($\log_{10}(g)$) are related to $L$, $R$ and $M$ according to Eqs.~\ref{eq:Teff} (Stefan-Boltzmann law) and \ref{eq:logg}, where $\sigma$ and $G$ are the Stefan-Boltzmann and gravitational constants respectively.

\begin{equation} \label{eq:Teff}
    T_\mathrm{eff}=\left(\frac{L}{4\pi\sigma R^2}\right)^{1/4}
\end{equation}

\begin{equation} \label{eq:logg}
    \log_{10}(g)=\log_{10}\left(\frac{GM}{R^2\ [1\ \mathrm{cm}\ \mathrm{s}^{-2}]}\right)
\end{equation}

\subsection{Atmosphere Modeling}

Model atmospheres were calculated with version 9 of the \texttt{ATLAS} code \citep{ATLAS5,ATLAS_Linux,ATLAS_howto,ATLAS}. The code attains high efficiency by sampling opacity from pre-tabulated opacity distribution functions (ODFs), described in \citet{ODFs} and \citet{Carbon}. The \texttt{ATLAS} suite also contains the \texttt{DFSYNTHE} program \citep{DFSYNHTE} that may be used to calculate ODFs for any given set of abundances, and the \texttt{SYNTHE} code \citep{SYNTHE} that computes the emergent spectrum from converged \texttt{ATLAS} models by sampling the opacity directly at the wavelengths of interest. All atmosphere models in this study were calculated at zero metallicity and with a more precise estimate of the primordial helium mass fraction, $Y=0.2448$, adopted from \citet{Valerdi}.

For this project, we developed a universal \texttt{Python} dispatcher that combines all three codes (originally written in \texttt{Fortran}) in a single user-friendly pipeline, complete with intermediate consistency checks and comprehensive documentation. Our dispatcher \citep{BasicATLAS} is available online\footnote{\href{https://github.com/Roman-UCSD/BasicATLAS}{\url{https://github.com/Roman-UCSD/BasicATLAS}}}.

\texttt{ATLAS} stratifies the atmosphere into $72$ plane-parallel layers spanning the range of Rosseland mean optical depths ($\tau$) from $\tau=10^2$ at the bottom to $\tau=10^{-7}$ at the top. For all models with stellar masses below $20\ M_\odot$, the ODFs were calculated following the ``new'' format \citep{new_ODFs} at $57$ temperatures between $10^{3.3}\ \mathrm{K}$ and $10^{5.3}\ \mathrm{K}$. At higher masses, the range fails to accommodate the deepest layers of the atmosphere that may exceed $10^{5.3}\ \mathrm{K}$ in temperature. As such, a second set of ODFs was calculated with an extended upper temperature limit of $10^{5.85}\ \mathrm{K}$. For those calculations, the definition of the temperature grid stored in the variable \texttt{TABT} of the \texttt{LINOP()} subroutine in the \texttt{ATLAS} source code was modified according to the altered ODF format. Furthermore, the wavelength grid for opacity and radiation field sampling in \texttt{ATLAS}, that by default spans from $\approx 9\ \mathrm{nm}$ to $\SI{160}{\micro\metre}$, had to be extended, first to $4\ \mathrm{nm}$ at stellar masses over $6\ M_\odot$, and then to $0.1\ \mathrm{nm}$ at stellar masses over $29\ M_\odot$. These extensions avoid errors in flux density and opacity integration due to significant contributions outside the default wavelength range. For each model, the adopted wavelength range was validated by ensuring that both the Planck function ($B_\nu(T)$) and its derivative ($dB_\nu/dT$) drop below $0.1\%$ of their maximum values at the wavelength range bounds in each layer of the atmosphere. The changes were applied to the \texttt{WBIG} variable in the \texttt{BLOCKR()} subroutine of \texttt{ATLAS}.

Atmosphere calculations in \texttt{ATLAS} are carried out through iterative improvements of an initial guess of the temperature profile throughout the atmosphere until a new profile is found that meets both the hydrostatic equilibrium condition for the prescribed surface gravity and the energy equilibrium condition for the prescribed effective temperature. On each iteration, the hydrostatic equilibrium condition is applied first to determine the pressure profile corresponding to the current temperature profile. Energy equilibrium is then evaluated throughout the atmosphere to determine corrections for the current temperature profile as well as the current percentage error in the flux and its derivative in each layer.

Since hydrostatic equilibrium is a hard requirement in \texttt{ATLAS}, no models can be calculated for stars above the Eddington limit -- a critical luminosity (or, equivalently, critical effective temperature) above which the radiation pressure gradient begins to exceed gravitational attraction. To determine this limit, we considered the range of gravities between $\log_{10}(g)=4.6$ and $\log_{10}(g)=5.2$ and searched for the maximum effective temperature ($T_\mathrm{eff}^{\mathrm{max}}$) at which \texttt{ATLAS} is able to find a solution that is both in hydrostatic equilibrium and has flux and flux derivative errors below $1\%$ over the course of $50$ iterations using the grey temperature profile \citep[Ch. 3]{integral_equation},

\begin{equation} \label{eq:grey}
    T(\tau)=T_\mathrm{eff}\left(\frac{3}{4}\tau + \frac{1}{2} \right)^{1/4}
\end{equation}

\noindent as the initial guess, where $T(\tau)$ is the temperature at optical depth $\tau$. The calculated values of $T_\mathrm{eff}^{\mathrm{max}}$ showed a nearly perfect exponential dependence on $\log_{10}(g)$ with deviations not exceeding $1\%$ throughout the entire range of considered gravities. The functional form of the relationship is given by

\begin{equation} \label{eq:3}
   \log_{10}(T_\mathrm{eff}^{\mathrm{max}})=C_1 \log_{10}(g) + C_2
\end{equation}

\noindent where $C_1$ and $C_2$ are the best-fit parameters calculated as $C_1=0.2516\pm 0.0013$ and $C_2=3.768\pm 0.006$ for CGS units.
Eq.~\ref{eq:3} is directly comparable to the commonly adopted ``classical'' Eddington limit (\citealt[Ch. 7-2]{integral_equation}, \citealt[Ch. 1]{eddington_limit_2}, \citealt{eddington_limit_1}) in CGS units, based on grey opacity dominated by Thomson scattering off electrons and local thermodynamic equilibrium (LTE):

\begin{equation} \label{eq:edd}
   \log_{10}(T_\mathrm{eff}^{\mathrm{max}})=\frac{1}{4} \log_{10}(g) + \frac{1}{4}\log_{10}\left(\frac{c \rho}{\sigma n_e \sigma_T}\right)
\end{equation}

Here, $c$ is the speed of light, $\rho$ is the mass density of the atmospheric layer, $n_e$ is the corresponding electron number density, and $\sigma_T$ is the Thomson scattering cross-section for an electron. When the layer is fully ionized, $n_e/\rho$ only depends on the helium mass fraction:

\begin{equation} \label{eq:ne}
   \frac{n_e}{\rho}=2\frac{Y}{m_\mathrm{He}} + \frac{1-Y}{m_\mathrm{H}}
\end{equation}

\noindent where $m_\mathrm{He}$ and $m_\mathrm{H}$ are the helium and hydrogen atomic masses respectively.
The ``classical'' equivalents of $C_1$ and $C_2$, denoted $C_1^{(T)}$ and $C_2^{(T)}$ respectively, can be evaluated numerically as $C_1^{(T)}=0.25$ and $C_2^{(T)}=3.795$. Since $C_1^{(T)}\approx C_1$ and $C_2^{(T)}>C_2$, the exact Eddington limit estimated with \texttt{ATLAS} is slightly lower than its ``classical'' counterpart at all considered surface gravities due to additional non-grey opacity sources and non-LTE effects in the atmosphere in the complete treatment.

\begin{figure}
    \centering
    \includegraphics[width=1\columnwidth]{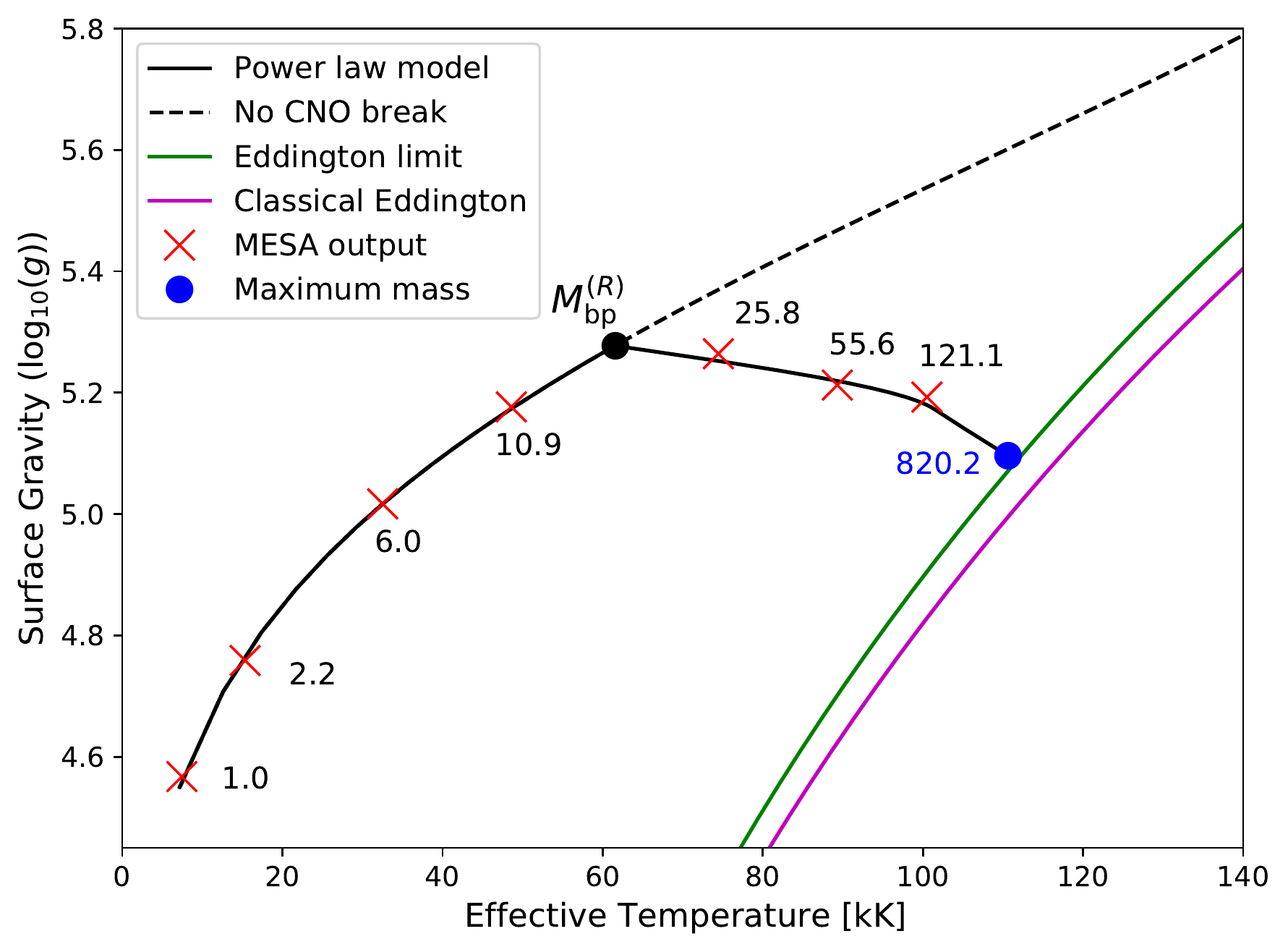}
    \caption{Surface parameters of ZAMS Population III stars compared to the exact calculation of the Eddington limit, including non-LTE effects and non-grey atmospheric opacity. The dashed curve traces out an alternative locus of surface parameters in the absence of the CNO break in the mass-radius relationship at $M_\mathrm{bp}^{(R)}\approx16\ M_\odot$. Surface parameters extracted directly from selected \texttt{MESA} models are shown with red markers and labelled by the initial masses (in $M_\odot$). The maximum stellar mass with a convergent \texttt{ATLAS} atmosphere is shown with the blue circle at $820.2\ M_{\odot}$. The ``classical'' (grey atmosphere, LTE) Eddington limit from Eq.~\ref{eq:edd} is shown for reference. Note that the exact Eddington limit results in a lower maximum mass value.}
    \label{fig:teffvslogg}
\end{figure}

Eq.~\ref{eq:3} is plotted in Fig.~\ref{fig:teffvslogg} alongside the locus of surface parameters of ZAMS Population III stars derived from our analytic mass-radius and mass-luminosity relationships in Eqs.~\ref{eq:1} and \ref{eq:2}. The intersection between the two curves approximately represents the maximum mass of Population III stars with atmospheres in hydrostatic equilibrium, which also serves as the maximum initial stellar mass considered in this study. The maximum mass was calculated as $820.2\ M_\odot$ by gradually increasing the initial stellar mass of the model in increments of $0.1\ M_\odot$ until no convergent atmosphere solution could be found. Figure~\ref{fig:teffvslogg} emphasizes the importance of the CNO cycle in Population III stars, as the shape of the surface parameters locus is clearly dominated by the power law break in the derived mass-radius relationship at $M\approx 16\ M_\odot$.

\begin{figure}
    \centering
    \includegraphics[width=1\columnwidth]{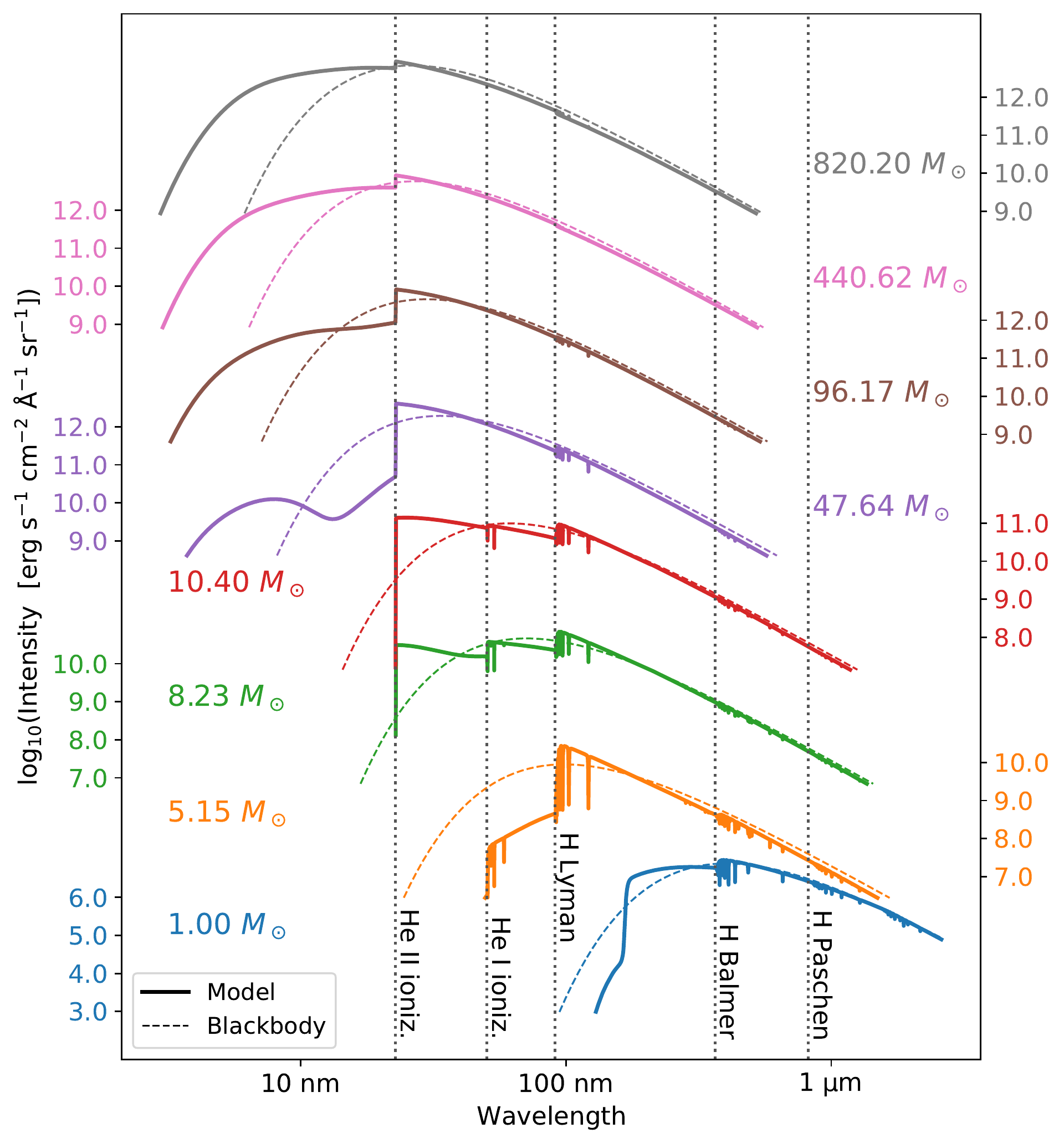}
    \caption{Synthetic spectra of ZAMS Population III stars for initial stellar masses between $1\ M_\odot$ and the Eddington limit ($820.2\ M_\odot$). For clarity, each spectrum is shown on a separate color-coded vertical scale. The blackbody spectra at the corresponding effective temperatures are shown in dashes for reference. Important bound-free absorption breaks are highlighted with vertical lines and labelled.}
    \label{fig:spectra}
\end{figure}

Overall, we calculated $59$ \texttt{ATLAS} atmospheres, logarithmically sampling the range of initial stellar masses between $1\ M_\odot$ and $820.2\ M_\odot$ and using the derived analytic relationships for surface parameters. For each model, the number of iterations was incremented in batches of $15$ until the maximum flux error and the maximum flux derivative error dropped below the standard convergence requirements of $1\%$ and $10\%$ respectively (\citealt{cookbook,convergence_standard}; see Appendix~\ref{sec:appendix} for details).
Synthetic spectra for each model were then calculated with \texttt{SYNTHE} between $0.5\ \mathrm{nm}$ and $\SI{2.6}{\micro\metre}$ at the resolution of $\lambda/\delta\lambda=6\times10^5$. The chosen wavelength range ensures that the flux density falls below $1\%$ of its maximum value at the range bounds for all calculated model atmospheres. To account for the limited buffer size in \texttt{SYNTHE}, the spectral synthesis for all atmospheres was carried out in three batches: between $0.5\ \mathrm{nm}$ and $14\ \mathrm{nm}$; between $14\ \mathrm{nm}$ and $400\ \mathrm{nm}$; and between $400\ \mathrm{nm}$ and $\SI{2.6}{\micro\metre}$. The calculations were run in parallel using the \textit{Triton Shared Computing Cluster} \citep{TSCC}. All calculated models are made public in our online repository\footnote{\href{https://atmos.ucsd.edu/}{https://atmos.ucsd.edu/}}. The key properties of all models as well as their convergence parameters are tabulated in Appendix~\ref{sec:appendix}. 

\begin{figure}
    \centering
    \includegraphics[width=1\columnwidth]{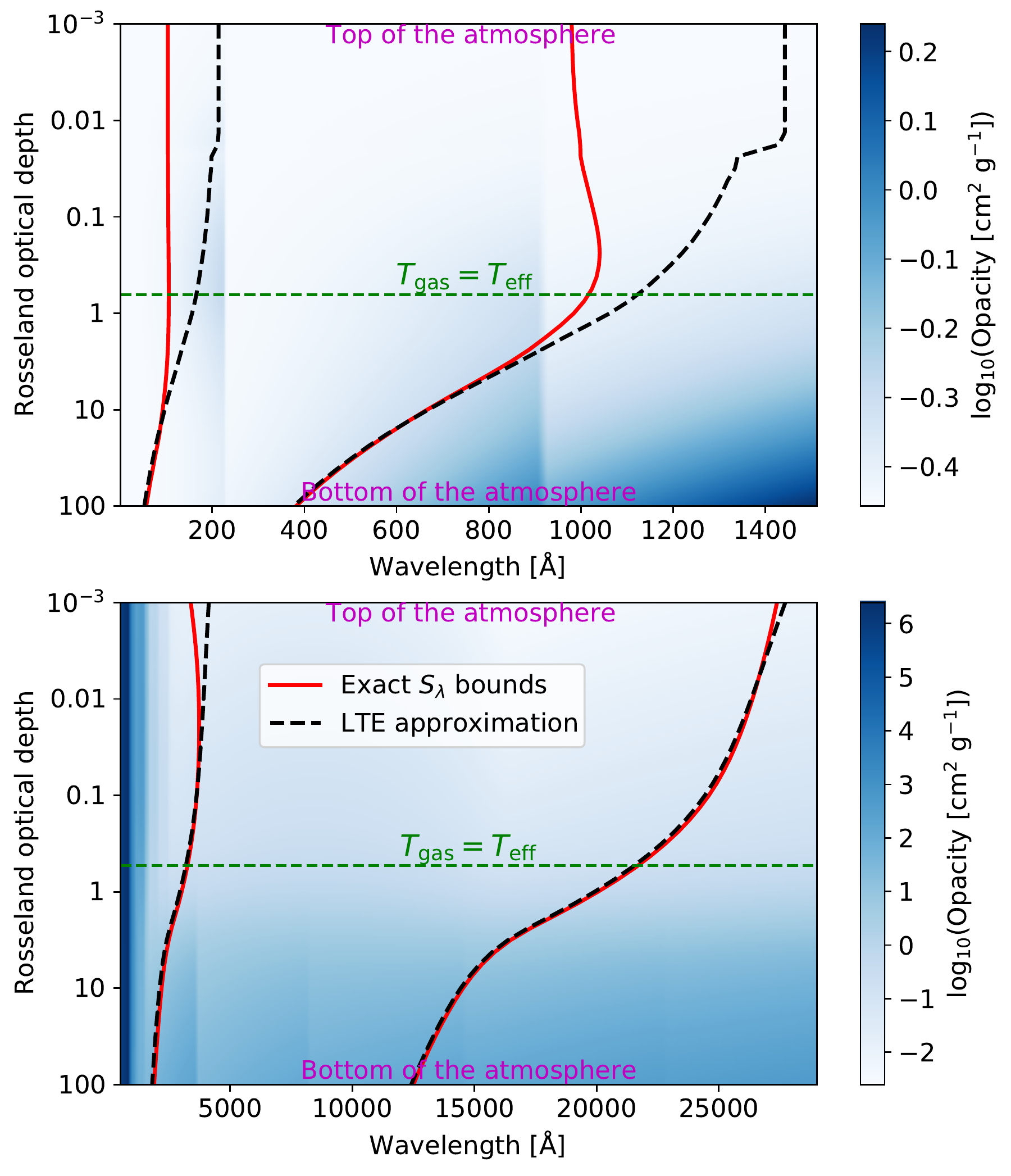}
    \caption{The $5^\mathrm{th}$ and the $95^\mathrm{th}$ percentiles of the wavelength distribution of the source function at a range of Rosseland mean optical depths for the solar atmosphere (\textit{Bottom}) and a near-Eddington Population III atmosphere with $M=820.2\ M_\odot$ (\textit{Top}). The cases of the exact solution for the source function and the LTE approximation are shown in solid red and dashed black respectively. The optical depth where the gas temperature matches the effective temperature of the star is highlighted for reference. The background color scheme corresponds to the total atmospheric opacity as a function of both wavelength and Rosseland optical depth. The discontinuous changes in opacity are due to bound-free absorption breaks.}
    \label{fig:snu}
\end{figure}

A few representative synthetic spectra are shown in Fig.~\ref{fig:spectra}. The spectra of stars with $M\gtrsim50\ M_\odot$ display a considerable flux excess blueward of the $\mathrm{He\ II}$ ionization break ($\approx 22.8\ \mathrm{nm}$) compared to their corresponding blackbody profiles. Since ultraviolet emission is heavily attenuated by the interstellar medium in the early universe, this effect results in an overall reduction of the observed brightness of Population III stars. The blue excess is primarily caused by non-LTE scattering of photons from deeper (and hotter) layers of the atmosphere. The departures from LTE in the radiation field are shown in Fig.~\ref{fig:snu} for the highest-mass Population III model considered in this study ($820.2\ M_\odot$) as well as a solar atmosphere model ($T_\mathrm{eff}=5770\ \mathrm{K}$, $\log_{10}(g)=4.44$, abundances from \citealt{roman_omegacen}). The figure shows the $5^\mathrm{th}$ and the $95^\mathrm{th}$ percentiles of the wavelength distribution of the source function ($S_\lambda$) for the case of LTE ($S_\lambda=B_\lambda$, no scattering) and the complete solution of the integral equation for $S_\lambda$ \citep[Ch. 6-1]{integral_equation}. For the solar model, both cases are nearly indistinguishable in all but the outermost layers of the atmosphere that do not contribute to the emergent spectrum significantly. On the other hand, the departure from LTE is far more prominent in the Population III atmosphere, with a noticeable blue excess in the radiation field at Rosseland mean optical depths shallower than $\sim 10$.

\reviewhighlight{The line features in Fig.~\ref{fig:spectra} diminish at higher masses due to the reduced populations of neutral species in the atmosphere required for bound-bound absorption. Selected lines may appear stronger than shown here due to unaccounted higher-order NLTE effects (e.g. overpopulation of excited levels as described in \citealt{NLTE_overpopulation}), as captured in model sets with a more detailed treatment of NLTE line profiles \citep{NLTE_models_1,NLTE_models_2,Rydberg_JWST}. However, the impact of narrow line features on broadband synthetic photometry is expected to be insignificant, especially in the JWST bands chosen in this study (see Section~\ref{sec:results}) that mostly occupy the comparatively line-free wavelength interval between the Lyman series of hydrogen and the Fowler series of ionized helium. To verify this claim, we recalculated our synthetic photometry (introduced in Section~\ref{sec:results}) with all line profiles artificially strengthened by the extreme factor of $10$. We found that at stellar masses over $100\ M_\odot$, the predicted magnitudes do not deviate from their nominal values by more than $0.003\ \mathrm{mag}$ in the chosen JWST bands.}

\section{Observable parameters} \label{sec:results}

\begin{figure*}
    \centering
    \includegraphics[width=1\textwidth]{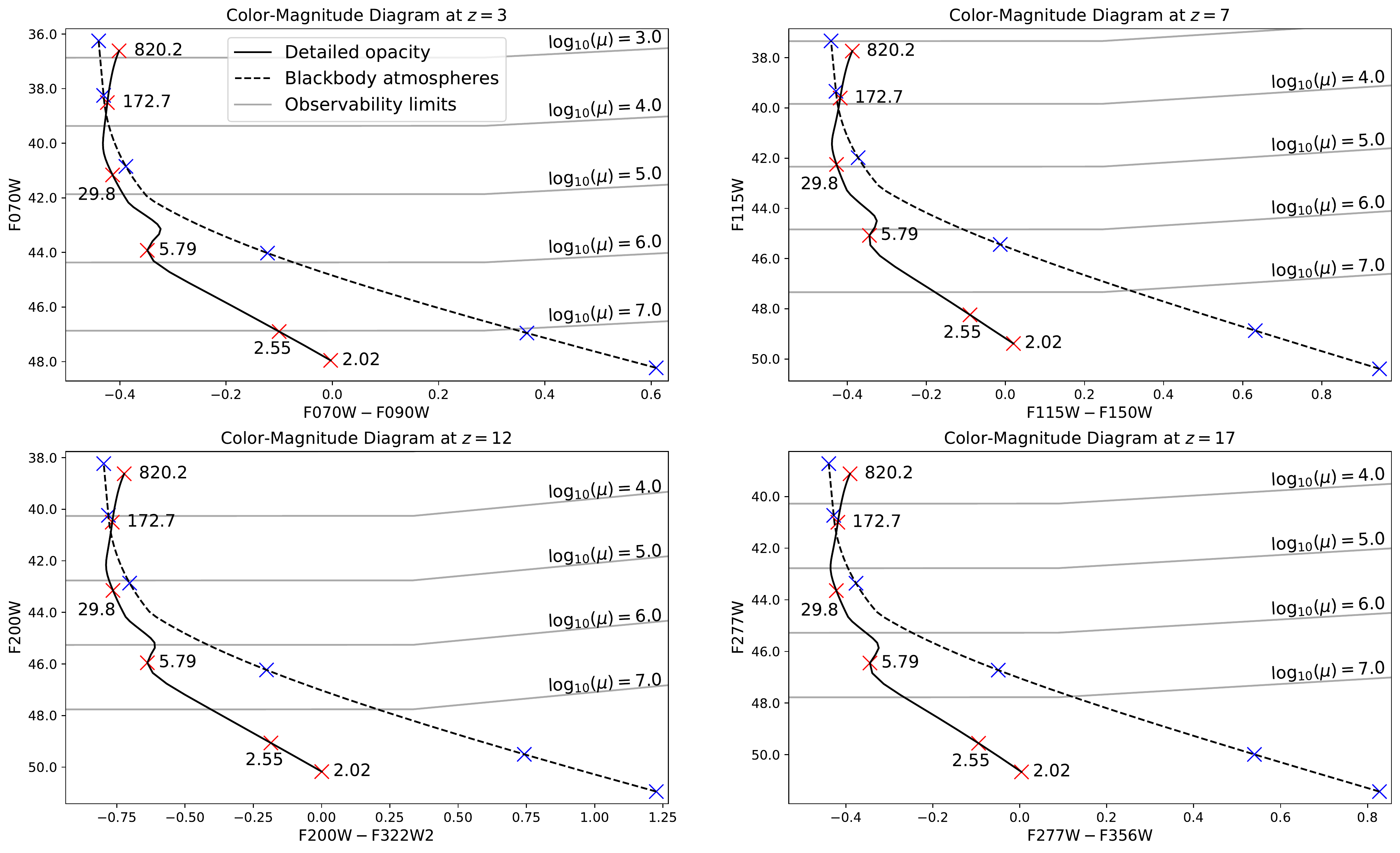}
    \caption{
    Color-magnitude diagrams of ZAMS Population III stars in the most optimal JWST bands listed in Table~\ref{tab:optimal_bands} at four different redshifts. Synthetic photometry is shown for both detailed opacity calculations based on model atmospheres and the corresponding blackbody profiles. Selected initial stellar masses are indicated in both cases with red and blue markers respectively. The red markers are labelled in solar masses. The blue markers correspond to the same masses as the red markers in the same order along the color-magnitude curve. Observability limits for JWST are shown in grey and labelled with the required gravitational lensing magnification, $\mu$.}
    \label{fig:colormag}
\end{figure*}

\begin{figure}
    \centering
    \includegraphics[width=1\columnwidth]{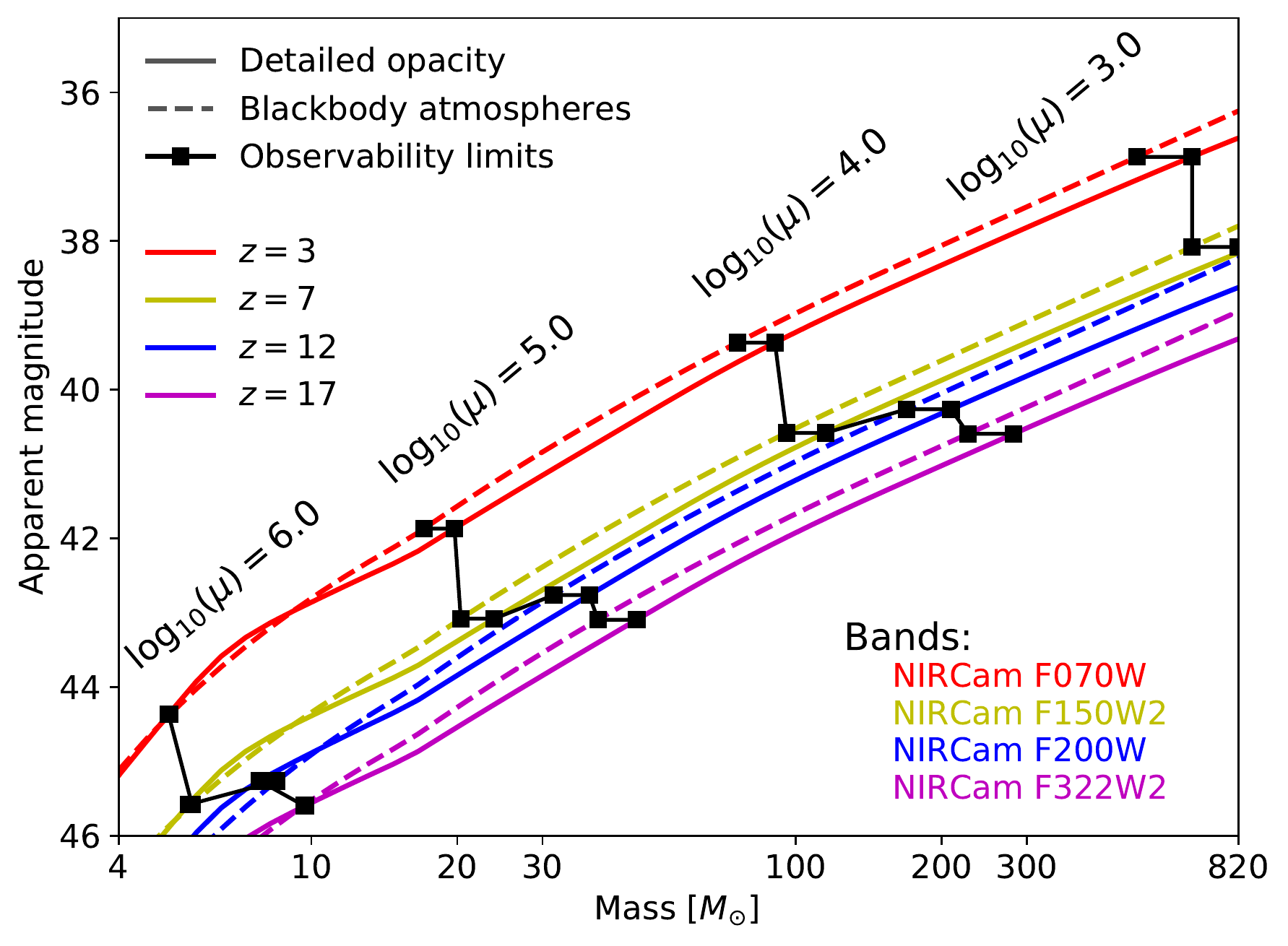}
    \caption{Mass-magnitude relationships for ZAMS Population III stars in the most optimal JWST bands listed in Table~\ref{tab:optimal_bands} at four different redshifts, color-coded. Note that a different band is used at each redshift. The equivalent relationships for the corresponding blackbody profiles are shown in dashed lines for comparison. The observability limits for JWST are indicated with black markers for each relationship shown. The displayed observability limits are grouped by the required gravitational lensing magnitification, $\mu$.}
    \label{fig:mass_luminosity}
\end{figure}

Color-magnitude diagrams (CMDs) in Fig.~\ref{fig:colormag} and mass-magnitude relationships in Fig.~\ref{fig:mass_luminosity} are provided for Population III stars in the most efficient JWST transmission bands. All synthetic photometry in this study was carried out in the \texttt{ABMAG} system \citep{ABmag} from the newly calculated synthetic spectra. For comparison, we also provide equivalent results for blackbody atmospheres at the corresponding effective temperatures.

\subsection{High-redshift synthetic photometry}
In \texttt{ABMAG}, the apparent magnitude, $m_\mathrm{AB}$, is calculated from the observed spectrum as:

\begin{equation} \label{eq:5}
    m_{AB} = -2.5 \log_{10}\left(\frac{\int (\nu)^{-1} f_{\nu}(\nu)e(\nu) d\nu} {\int 3631\ \mathrm{Jy}\ (\nu)^{-1}e(\nu) d\nu}\right)
\end{equation}

Here, $f_\nu(\nu)$ is the apparent flux density per unit frequency, $e(\nu)$ is the efficiency of the instrument, and the integrals are evaluated over all frequencies ($\nu$) in the frame of reference of the observer. The factor of $(\nu)^{-1}$ is included to adapt the relationship to a photon-counting instrument \citep{photon_counting}. Equivalently, Eq.~\ref{eq:5} may be written in terms of the observed wavelength, $\lambda$, to match the output of \texttt{SYNTHE}:


\begin{equation} \label{eq:6}
    m_{AB} = -2.5 \log_{10}\left(\frac{\int \lambda f_{\lambda}(\lambda) e(\lambda) d\lambda} {\int 3631\ \mathrm{Jy}\ (\lambda)^{-1} e(\lambda) c d\lambda}\right)
\end{equation}

At high redshift, the apparent flux density, $f_\lambda$, in Eq.~\ref{eq:5} and Eq.~\ref{eq:6} is derived from the modelled surface flux density, $F_\lambda$:


\begin{equation} \label{eq:7}
    f_\lambda(\lambda)=F_\lambda(\lambda_\mathrm{e})T(\lambda,z)\frac{R^2}{(1+z)D_L^2}
\end{equation}

\noindent (e.g. see \citealt{high_z_photometry}), where $\lambda_\mathrm{e}=\lambda/(1+z)$ is the emitted wavelength at redshift $z$, $R$ is the radius of the star and $D_L$ is the luminosity distance to the star. In the equation, $T(\lambda,z)$ is the integrated transmissivity of the interstellar medium across the line of sight to the source. In the wavelength range of interest, the most significant contributions to interstellar attenuation are the bound-free and bound-bound absorption by ground state neutral hydrogen at $\lambda_\mathrm{e}\leq \SI{1215.67}{\angstrom}$ -- the Lyman $\alpha$ wavelength.

\reviewhighlight{The photons absorbed by the interstellar medium will be re-emitted at longer wavelengths. Depending on the dynamical evolution of the medium under radiative feedback and the spatial density of Population III stars, this reprocessed radiation may contribute significantly to the observed spectrum of the star or have low impact on direct observations due to dilution over large surface areas (e.g. compare ``type A'' and ``type C'' environments in \citealt{ISM_reprocessing}; also see \citealt{dynamic_ISM}). In this study, we focus on the purely stellar spectra; however, see \citet{Rydberg_JWST,ISM_modelling_1,ISM_modelling_2,ISM_modelling_3} for various approaches to detailed feedback modelling.}

At $z\leq 7$, we adopt $T(\lambda,z)$ from the numerical simulation in \citet{Meiksin}. At higher redshifts, the absorption by neutral hydrogen is sufficiently strong to be well-approximated by a hard cut-off:

\begin{equation} \label{eq:cutoff}
    T(\lambda,z;z>7)\approx\begin{cases}
        0,& \text{if } \lambda \leq (1+z)\ \SI{1215.67}{\angstrom}\\
        1,& \mathrm{otherwise} \\
    \end{cases}
\end{equation}

As will be demonstrated, even at redshifts below $z=7$ considered in this study, the most appropriate JWST bands for detecting Population III stars have blue cut-offs at $\lambda_\mathrm{e}\gtrsim \SI{1215.67}{\angstrom}$, thereby ensuring that Eq.~\ref{eq:cutoff} remains a good approximation at all considered redshifts.

The luminosity distance, $D_L$, is calculated as a function of redshift as:

\begin{equation} \label{eq:9}
    D_L = (1+z)c \int_0^z {\frac{1}{H(z')}dz'}
\end{equation}

\noindent \citep[Ch. 2.2]{dodelson} where $H(z')$ is the Hubble parameter \citep[Ch. 2.4]{dodelson}: 

\begin{equation} \label{eq:10}
    H^2(z) = H_0^2\left(\Omega_r(1+z)^4 + \Omega_M(1+z)^3 + \Omega_\Lambda\right)
\end{equation}

In Eq.~\ref{eq:10}, $H_0$ is the Hubble constant and $\Omega_r$, $\Omega_M$ and $\Omega_\Lambda$ are the fractional present-day contributions of radiation (including relativistic matter), non-relativistic matter, and dark energy, respectively. We adopt the Hubble constant value of $H_0=67.4\ \mathrm{km}\ \mathrm{s}^{-1}\ \mathrm{Mpc}^{-1}$ and $\Omega_M = 0.315$ in accordance with \citet{H0}. The universe is assumed to be flat ($\Omega_\Lambda = 1 - \Omega_r - \Omega_M$) and $\Omega_r$ is calculated as \citep[Ch. 2.4.4]{dodelson}:

\begin{equation} \label{eq:OmR}
    \Omega_r = \left(1+\frac{7}{8}\left(\frac{4}{11}\right)^{\frac{4}{3}}N_\mathrm{eff}\right) \frac{4\sigma}{c^3} \frac{T^4_\mathrm{CMB}}{\rho_\mathrm{crit}}
\end{equation}

We adopt $N_\mathrm{eff} = 3.04$ \citep{Neff} as the effective number of neutrino flavors, $T_\mathrm{CMB} = 2.725\ \mathrm{K}$ \citep{CMB} as the present day temperature of the cosmic microwave background, and $\rho_\mathrm{crit} = 3/(8\pi G)H_0^2$ as the critical density of the universe. The calculation implicitly assumes massless neutrinos. The accuracy of these cosmological assumptions is examined in Section~\ref{sec:cosmo}. 

\subsection{Choice of bands}

This study considers detecting Population III stars with JWST at $z\in(3,7,12,17)$. The chosen range spans between the redshift of the predicted saturation of the Population III star formation rate in \citet{popIII_high_z} and the redshift of the candidate Population III ionization source in \citet{Fosbury}. If the aim of the experiment is a simple detection of a Population III candidate in a single band, the optimal observation band for each stellar mass and redshift may be chosen by seeking the largest value of the predicted signal-to-noise ratio. At the lowest stellar masses, this condition will be met by a wide band, situated closest to the peak wavelength of the model spectrum. Due to extensive attenuation of flux by the interstellar medium at wavelengths shorter than the Lyman $\alpha$, the optimal band at higher masses remains redward of the Lyman $\alpha$ in the observer's frame of reference instead of following the peak wavelength. The transition occurs around $\sim10\ M_\odot$ for $z=3$ and at $\lesssim 3\ M_\odot$ for $z\geq 7$. Since high-mass Population III stars are overwhelmingly more likely to be observable, the same high-mass optimal band may be safely employed for all Population III candidates at a given redshift.

We calculated the limiting magnitudes in all JWST Near Infrared Camera (NIRCam) and Mid-Infrared Instrument (MIRI) bands using the JWST Exposure Time Calculator \citep{JWST_ETC} as the faintest magnitudes resulting in a signal-to-noise ratio of $3$ in a $\sim10\ \mathrm{hr}$ exposure (NIRCam: 20 groups, 9 integrations, \texttt{DEEP2} readout pattern; MIRI: 100 groups, 132 integrations, \texttt{FASTR1} readout pattern) for a flat frequency continuum.
The best band for each redshift was chosen as the one corresponding to the smallest difference between the expected apparent magnitude of Population III stars in the high-mass regime and the limiting magnitude of the band. The chosen bands are listed in the ``\textit{Best single}'' column of Table~\ref{tab:optimal_bands}.

\begin{deluxetable}{l|c|ll}
\tablenum{2}
\tablecaption{Optimal bands for Population III detection\label{tab:optimal_bands}}
\tablewidth{\columnwidth}
\tablehead{
\colhead{$z$} & \colhead{Best single} & \colhead{Best pair} & \colhead{} 
}
\startdata
   $3$ & NIRCam \texttt{F070W} & NIRCam \texttt{F070W} & NIRCam \texttt{F090W} \\
   $7$ & NIRCam \texttt{F150W2} & NIRCam \texttt{F115W} & NIRCam \texttt{F150W} \\
   $12$ & NIRCam \texttt{F200W} & NIRCam \texttt{F200W} & NIRCam \texttt{F322W2} \\
   $17$ & NIRCam \texttt{F322W2} & NIRCam \texttt{F277W} & NIRCam \texttt{F356W} \\
\enddata
\end{deluxetable}

\begin{deluxetable}{lr|lr}
\tablenum{3}
\tablecaption{Limiting magnitudes\label{tab:lim_mag}}
\tablewidth{\columnwidth}
\tablehead{
Band & Lim. mag & Band & Lim. mag
}
\startdata
   NIRCam \texttt{F070W} & $29.3664$ & NIRCam \texttt{F090W} & $29.6480$ \\
   NIRCam \texttt{F150W2} & $30.5796$ & NIRCam \texttt{F115W} & $29.8404$ \\
   NIRCam \texttt{F150W} & $30.0816$ & NIRCam \texttt{F200W} & $30.2616$ \\
   NIRCam \texttt{F322W2} & $30.5926$ & NIRCam \texttt{F277W} & $30.2774$ \\
   NIRCam \texttt{F356W} & $30.3625$ &  &  \\
\enddata
\caption{Calculated limiting magnitudes are listed for JWST bands in Table~\ref{tab:optimal_bands}, assuming the detection signal-to-noise ratio of $3$ in a $\sim10\ \mathrm{hr}$ exposure (20 groups, 9 integrations, \texttt{DEEP2} readout pattern) of a flat frequency continuum.}
\end{deluxetable}

A more detailed experiment may be designed with the aim of measuring the colors of Population III candidates in addition to simple detection, requiring a choice of two filters without significant overlap in their transmission profiles. We determine the optimal pairs of JWST filters for each redshift by considering all possible non-overlapping pairs of bands and choosing the one with the smallest \textit{average} difference between the predicted magnitude of Population III candidates and the limiting magnitude of the band. As before, the choices are made in the high-mass regime. The resulting optimal pairs of filters are listed in the ``\textit{Best pair}'' column of Table~\ref{tab:optimal_bands}.

NIRCam bands were found to be most optimal for both simple detection and color measurements. The limiting magnitudes of all bands chosen in Table~\ref{tab:optimal_bands} are listed in Table~\ref{tab:lim_mag}.

\subsection{Results}

Predicted color-magnitude diagrams are presented in Fig.~\ref{fig:colormag} for all four redshifts in the most optimal JWST band pairs listed in Table~\ref{tab:optimal_bands}. Synthetic photometry for blackbody atmospheres at the same range of effective temperatures is shown in the figure as well for comparison. The overall trend of the color-magnitude relationship is nearly unchanged between redshifts, as the most efficient bands are placed in similar positions with respect to the redshifted energy density distribution predicted by the model atmospheres. At low masses ($M\lesssim100\ M_\odot$), predicted colors shift blueward with increasing effective temperature, with the exception of a brief inversion of the trend around $6\ M_\odot$. We refer to this color-magnitude diagram feature as the ``\textit{helium loop}'', as the inversion is caused by bound-free absorption of singly ionized helium in the second excited state (``Fowler break'', $\lambda_\mathrm{e}=205.1\ \mathrm{nm}$, \citealt{fowler_break}). Once formed, the break disproportionately suppresses flux in the bluer band, resulting in the redder overall color.

In the high-mass regime ($M\gtrsim100\ M_\odot$), the predicted color shifts redward with increasing temperature due to progressively decreasing contributions of free-free and bound-free opacities, both of which vary as $\lambda^3$ and, therefore, redistribute the flux towards shorter wavelengths. At high masses, Population III stars are predicted to be slightly fainter than blackbodies with identical effective temperatures due to the non-LTE distribution of the radiation field illustrated in Fig.~\ref{fig:snu}.

Magnitude predictions for the most efficient single-band observations at each redshift are shown in Fig.~\ref{fig:mass_luminosity} as functions of mass. Both Figs.~\ref{fig:colormag} and \ref{fig:mass_luminosity} also contain the estimated JWST observability limits for different gravitational lensing magnifications, from $\mu=10^3$ (approximate minimum required magnification for direct observations of Population III stars) to $\mu=10^7$ (maximum theoretical magnification from \citealt{max_magnification}). The observability limits are based on the calculated limiting magnitudes in each band listed in Table~\ref{tab:lim_mag}.

\subsection{Cosmological parameters} \label{sec:cosmo}

\begin{deluxetable}{cccccc}
\tablenum{4}
\tablecaption{Effect of cosmological parameters\label{tab:cosmo}}
\tablewidth{\columnwidth}
\tablehead{
\multirow{2}{*}{\begin{tabular}{c}\end{tabular}} & \multirow{2}{*}{\begin{tabular}{c}Adopted\\Value\end{tabular}} & \multicolumn{2}{c}{Range} &  \multicolumn{2}{c}{Effect $\mathrm{[mag]}$} \\
 & & Min & Max & Min & Max
}
\startdata
   $H_0$\tablenotemark{a} & $67.4$ & $66.9$ & $75.0$ & $+0.02$ & $-0.23$ \\
   $\Omega_M$ & $0.315$ & $0.300$ & $0.322$ & $+0.04$ & $-0.02$ \\
   $m_\nu$\tablenotemark{b} & $0.0$ & $0.0$ & $0.9$ & $0.0$ & $-0.15$ \\
\enddata
\tablenotetext{a}{Hubble constant in $\mathrm{km}\ \mathrm{s}^{-1}\ \mathrm{Mpc}^{-1}$}
\tablenotetext{b}{Neutrino mass in $\mathrm{eV}\ c^{-2}$, assumed identical for all neutrino species}
\end{deluxetable}

Predicted colors and magnitudes of Population III stars depend on the cosmological parameters adopted when calculating the luminosity distance, $D_L$ (Eq.~\ref{eq:9}). Since $D_L$ is independent of the physical properties of the star, the effect is identical for all initial masses. In this section, we estimate the magnitude of the effect for individual variations in the Hubble constant, $H_0$; the present-day matter contribution, $\Omega_M$ and the average neutrino mass, $m_\nu$. The nominal value of each parameter adopted in this study, the considered range of variation, and the shift in the predicted magnitudes at the lower and upper bounds of the considered range are provided in Table~\ref{tab:cosmo}. All tests are carried out at the largest considered redshift, $z=17$, where the effect is expected to be most significant.

For the Hubble constant, the adopted range spans from the lower error bound of the adopted nominal value ($67.4\pm0.5\ \mathrm{km}\ \mathrm{s}^{-1}\ \mathrm{Mpc}^{-1}$, \citealt{H0}), based on Planck observations of the cosmic microwave background (CMB), to the upper error bound of the local Hubble constant estimate in \citet{HubbleConst} ($73.24\pm1.74\ \mathrm{km}\ \mathrm{s}^{-1}\ \mathrm{Mpc}^{-1}$), based on the updated distance calibration to type Ia supernovae.

The discrepancy between the CMB and local measurements of $H_0$, known as the \textit{Hubble Tension}, depends on the adopted $\Omega_M$ prior. The two measurements have been shown to be consistent at $95\%$ confidence for $\Omega_M\lesssim 0.3$ \citep{hubble_tension}. We therefore adopt $\Omega_M=0.3$ as the lower bound on the variation range and take the error in the nominal value ($0.315\pm0.007$, \citealt{H0}) as the upper bound.

To estimate the effect of massive neutrinos, we replace the relativistic neutrino density in Eq.~\ref{eq:OmR} with the approximation for neutrinos with identical masses in \citet{neutrino_formula}, implemented in \texttt{Astropy} (\citealt{astropy_1,astropy_2}). We adopted $m_\nu=0.9\ \mathrm{eV}\ c^{-2}$ as the maximum neutrino mass, measured in the Karlsruhe Tritium Neutrino (KATRIN) experiment \citep{NeutrinoMass}.

As demonstrated in Table~\ref{tab:cosmo}, $H_0$ and $m_\nu$ have the largest effect on the predicted photometry. Larger values for both parameters lead to shorter lookback times to a given redshift and, therefore, brighter apparent magnitudes of Population III stars. However, the gain in magnitude for both parameters was calculated to fall below $0.25$ at the most extreme, which is expected to remain within the measurement uncertainty at the adopted limiting magnitude signal-to-noise ratio of $3$.

\section{Conclusion} \label{sec:conclusion}

In this study, we calculated new evolutionary models and model atmospheres for ZAMS Population III stars in hydrostatic equilibrium. The new models were used to investigate the physical properties of the first stars in the universe as well as to produce predictions of their colors and magnitudes as may be observed in the near future with JWST under strong gravitational lensing. The analysis was carried out at a broad range of plausible redshifts for Population III stars from $z=3$ to $z=17$. Our predictions of Population III color-magnitude diagrams and mass-magnitude relations are provided in Figs.~\ref{fig:colormag} and \ref{fig:mass_luminosity} respectively. All predictions are given for the optimal JWST bands listed in Table~\ref{tab:optimal_bands}. Our other findings are listed below:

\begin{itemize}
    \item The mass-radius relationship of ZAMS Population III stars is well approximated by a broken power law, similar to their Population I counterparts. However, the break in the power law occurs at a much higher mass ($\approx 16\ M_\odot$ for Population III stars as opposed to $\approx 1\ M_\odot$ for Population I) due to the suppressed CNO cycle.
    \item Despite the initial absence of metals in Population III stars, the required amount of carbon to sustain the CNO cycle is produced at sufficiently high masses. The CNO cycle becomes the dominant energy production mechanism in Population III stars around $M\approx 20.5\ M_\odot$, in agreement with \citet{Yoon}.
    \item The mass-luminosity relationship of ZAMS Population III stars may be approximated as a power law with a variable power index that decreases at higher masses. This behavior is observed in Population I stars as well and is approximately consistent with the Eddington standard model.
    \item The evolution of true metal-free stars is nearly indistinguishable from the evolution of extremely metal-poor stars with $Z\lesssim 10^{-9}$. This result is more conservative but consistent with the $Z=10^{-8}$ limit derived in \citet{Windhorst}. Furthermore, both values agree with the lower bound of the expected threshold of the Population III / Population II transition ($Z_\mathrm{cr}\gtrsim 10^{-8}$; \citealt{crit_Z_1,crit_Z_2,crit_Z_3}).
    \item The maximum mass of ZAMS Population III stars, at which hydrostatic equilibrium is possible in the atmosphere (the Eddington limit) was calculated as $\approx 820\ M_\odot$. This value is well above the commonly considered range of initial masses for primordial stars, suggesting that hydrostatic equilibrium may be an adequate approximation in Population III models. The exact Eddington limit was found to be slightly lower than predicted by the classical formula (Eq.~\ref{eq:edd}) due to non-grey opacity sources in the atmosphere as well as non-LTE effects. The influence of the CNO cycle on the internal structure of Population III stars was determined to be a key factor in setting the maximum mass.
    \item Atmospheres of high-mass Population III stars host strongly non-LTE radiation fields, resulting in significant excess in the UV flux compared to the corresponding blackbody profiles.
    \item The color-magnitude diagrams of Population III stars depend strongly on the non-grey opacity sources in the atmosphere with notable features including the ``\textit{helium loop}'' at $M\sim 6\ M_\odot$ and the color-temperature inversion at $M\gtrsim 100\ M_\odot$. In general, ZAMS Population III stars are fainter than expected from blackbody profiles.
    \item At the lowest redshift ($z=3$), the highest-mass Population III stars considered in this study ($M\gtrsim 700\ M_\odot$) are just observable with a gravitational lensing magnification of $\mu\sim10^3$. A more plausible range of stellar masses ($M\gtrsim 100\ M_\odot$) would likely require $\mu\sim 10^4$. Such magnification is comparable to that inferred from previous detections of the most distant individual stars known (e.g. \citealt{even_more_distant_star}). At higher redshifts, the required magnification for an equivalent detection increases to $\mu\sim 10^5$.
    \item Our predictions of Population III observability do not depend significantly on the adopted cosmological parameters; however, the maximum calculated effect of $\sim 0.25\ \mathrm{mag}$ is comparable to the adopted signal-to-noise ratio of JWST observations and may therefore be measurable under more generous gravitational lensing conditions than the minimum detection requirement considered in this study.
\end{itemize}

This study is limited to ZAMS Population III stars and may therefore be considered a lower limit on true JWST observability since later evolution stages are generally expected to be more luminous and less attenuated by the interstellar medium. The reduction in surface gravity during the post-main sequence evolution of Population III stars may drive the highest-mass stars considered in this study above the calculated Eddington limit, requiring a more detailed modelling approach allowing for mechanical motion in the atmosphere as well as mass loss.

\reviewhighlight{In our predictions of the observational signatures of Population III stars, only radiation emitted directly from the stellar photospheres was considered. Realistic regions of Population III formation are likely to display significant flux contributions from the surrounding interstellar nebula. A follow-up study could deploy an analytic ionization model, as in \citet{ISM_modelling_1}, or a numerical simulation of radiative feedback, to derive the necessary corrections. In this context, the predictions drawn here may once again be interpreted as lower limits of the true observability of individual Population III sources.}

While the detailed modelling of atmospheric opacity was shown to produce noticeably different results from the commonly adopted blackbody approximation (e.g. \citealt{Windhorst,Fosbury}), the quantitative difference in predictions of the two approaches will likely remain within the measurement uncertainty for Population III candidates at the observability threshold (e.g. at $z=3$, the magnitude difference between the two approaches is $\approx 0.2\ \mathrm{mag}$ at the highest considered mass). However, the discrepancy may be detectable under marginally stronger gravitational lensing and should therefore be taken into account in more detailed observational studies of Population III stars.

Our overall result generally agrees with previous studies of Population III observability (e.g. \citealt{Windhorst,Rydberg_JWST}) that detection of the first stars in the universe may be possible with JWST under strong but realistic gravitational lensing, assuming sufficiently high stellar mass. Placing more specific constraints on the expected rate of detection remains challenging due to the highly debated initial mass function of Population III stars.

\reviewhighlight{Finally, we note that Population III stars likely formed in clusters or galaxies rather than in isolation (e.g. \citealt{Jaacks,ISM_reprocessing,pop_III_clusters_1,pop_III_clusters_2,pop_III_clusters_3,pop_III_clusters_4}). The combined luminosity of such objects makes them more accessible targets, requiring less extreme gravitational lensing. However, modelling the spectral energy distributions of Population III clusters is further complicated by the dependency on the highly uncertain initial mass function and the concurrent formation of Population III and Population II stars \citep{popIII_high_z,popIII_high_z_1,popIII_high_z_2,Jaacks} at later epochs.}

\vspace{5em}
M.\ Larkin acknowledges funding support from the University of California at San Diego (UCSD) Department of Physical Sciences Summer Research Award and the UCSD Triton Research \& Experiential Learning Scholars (TRELS) program.
R.\ Gerasimov acknowledges funding support from HST Program GO-15096, provided by NASA through a grant from the Space Telescope Science Institute, which is operated by the Association of Universities for Research in Astronomy, Incorporated, under NASA contract NAS5-26555.
This work was conducted at UCSD, which was built on the unceded territory of the Kumeyaay Nation.  Today, the Kumeyaay people continue to maintain their political sovereignty and cultural traditions as vital members of the San Diego community.  We acknowledge their tremendous contributions to our region and thank them for their stewardship. 

%

\facilities{JWST (NIRCam, MIRI)}


\software{
\texttt{Astropy} \citep{astropy_1,astropy_2},  
\texttt{Matplotlib} \citep{matplotlib},
\texttt{NumPy} \citep{numpy}, 
\texttt{SciPy} \citep{scipy},
\texttt{ATLAS} \citep{ATLAS5,ATLAS_Linux,ATLAS_howto,ATLAS},
\texttt{DFSYNTHE} \citep{DFSYNHTE},
\texttt{SYNTHE} \citep{SYNTHE},
\texttt{MESA} \citep{MESA,MESA_2,MESA_3,MESA_4,MESA_5}
}

\clearpage

\appendix

\section{Model parameters} \label{sec:appendix}

Table~\ref{tab:params} lists the defining parameters and convergence criteria of all \texttt{ATLAS-9} model atmospheres calculated in this study. Initial masses are sampled logarithmically between $1\ M_\odot$ and the estimated Eddington limit ($820\ M_\odot$). The corresponding stellar radii, luminosities, effective temperatures and surface gravities are calculated using Eqs.~\ref{eq:1}, \ref{eq:2}, \ref{eq:Teff} and \ref{eq:logg} as well as the best-fit parameters in Table~\ref{tab:best_fit}. Only the effective temperature and surface gravity are used as inputs to model atmospheres. The final convergence is parameterized in terms of the maximum absolute flux error and the maximum absolute flux derivative error with respect to the depth-integrated mass density (see \citealt{ATLAS5}). Flux errors and flux derivative errors are used to calculate the temperature corrections between iterations using the \citet{T_corr_1} scheme at large optical depths and the $\Lambda$-iteration scheme (\citealt{T_corr_2}, \citealt[Ch. 3-3]{integral_equation}) at shallow optical depths, respectively. All models calculated in this study meet the standard convergence target of flux error below $1\%$ and flux derivative error below $10\%$ \citep{cookbook,convergence_standard}.

\startlongtable
\begin{deluxetable}{ccccccc}
\tablenum{5}
\tablecaption{Parameters of Population III models calculated in this study\label{tab:params}}
\tablewidth{\textwidth}
\tablehead{
\colhead{Initial Mass} & \colhead{Effective Temperature} & \colhead{Surface gravity} & \colhead{Radius} & \colhead{Luminosity} & \colhead{Max Flux} & \colhead{Max Flux}\\
\colhead{$M\ [M_\odot]$} & \colhead{$T_\mathrm{eff}\ [\mathrm{K}]$} & \colhead{$\log_{10}(g)$} & \colhead{$R\ [R_\odot]$} & \colhead{$\log_{10}(L/L_\odot)$} & \colhead{Error $[\%]$} & \colhead{Derivative Error $[\%]$}
}
\startdata
$1.000$ & $7180$ & $4.550$ & $0.879$ & $0.267$ & $0.24$ & $6.58$ \\
$1.124$ & $8047$ & $4.581$ & $0.900$ & $0.485$ & $0.62$ & $1.54$ \\
$1.264$ & $9000$ & $4.612$ & $0.921$ & $0.700$ & $0.61$ & $3.11$ \\
$1.421$ & $10045$ & $4.642$ & $0.942$ & $0.911$ & $0.71$ & $7.49$ \\
$1.597$ & $11189$ & $4.673$ & $0.964$ & $1.118$ & $0.76$ & $8.30$ \\
$1.796$ & $12437$ & $4.704$ & $0.987$ & $1.322$ & $0.78$ & $7.88$ \\
$2.019$ & $13796$ & $4.734$ & $1.010$ & $1.523$ & $0.85$ & $6.84$ \\
$2.270$ & $15273$ & $4.765$ & $1.034$ & $1.719$ & $0.79$ & $5.00$ \\
$2.551$ & $16873$ & $4.796$ & $1.058$ & $1.912$ & $0.84$ & $3.89$ \\
$2.868$ & $18602$ & $4.827$ & $1.083$ & $2.102$ & $0.74$ & $4.94$ \\
$3.225$ & $20466$ & $4.857$ & $1.108$ & $2.288$ & $0.58$ & $5.76$ \\
$3.625$ & $22472$ & $4.888$ & $1.134$ & $2.471$ & $0.53$ & $5.73$ \\
$4.075$ & $24623$ & $4.919$ & $1.161$ & $2.650$ & $0.45$ & $5.17$ \\
$4.582$ & $26924$ & $4.949$ & $1.188$ & $2.825$ & $0.35$ & $4.50$ \\
$5.151$ & $29381$ & $4.980$ & $1.216$ & $2.997$ & $0.34$ & $3.23$ \\
$5.790$ & $31996$ & $5.011$ & $1.245$ & $3.165$ & $0.31$ & $1.63$ \\
$6.510$ & $34772$ & $5.042$ & $1.274$ & $3.330$ & $0.23$ & $1.31$ \\
$7.318$ & $37712$ & $5.072$ & $1.304$ & $3.491$ & $0.24$ & $0.93$ \\
$8.227$ & $40817$ & $5.103$ & $1.334$ & $3.648$ & $0.28$ & $0.60$ \\
$9.249$ & $44087$ & $5.134$ & $1.365$ & $3.802$ & $0.31$ & $0.65$ \\
$10.398$ & $47521$ & $5.164$ & $1.397$ & $3.953$ & $0.32$ & $0.47$ \\
$11.690$ & $51118$ & $5.195$ & $1.430$ & $4.100$ & $0.32$ & $0.36$ \\
$13.141$ & $54874$ & $5.226$ & $1.464$ & $4.243$ & $0.27$ & $0.27$ \\
$14.774$ & $58785$ & $5.256$ & $1.498$ & $4.383$ & $0.24$ & $0.23$ \\
$16.609$ & $62432$ & $5.276$ & $1.554$ & $4.519$ & $0.20$ & $0.26$ \\
$18.672$ & $65238$ & $5.270$ & $1.658$ & $4.652$ & $0.18$ & $0.31$ \\
$20.991$ & $68031$ & $5.265$ & $1.769$ & $4.781$ & $0.17$ & $0.46$ \\
$23.598$ & $70799$ & $5.260$ & $1.887$ & $4.906$ & $0.15$ & $0.49$ \\
$26.529$ & $73528$ & $5.254$ & $2.013$ & $5.028$ & $0.12$ & $0.56$ \\
$29.825$ & $76205$ & $5.249$ & $2.148$ & $5.147$ & $0.14$ & $0.59$ \\
$33.529$ & $78819$ & $5.243$ & $2.291$ & $5.261$ & $0.15$ & $0.70$ \\
$37.694$ & $81355$ & $5.238$ & $2.445$ & $5.373$ & $0.21$ & $0.74$ \\
$42.376$ & $83800$ & $5.233$ & $2.608$ & $5.480$ & $0.26$ & $0.75$ \\
$47.639$ & $86143$ & $5.227$ & $2.783$ & $5.584$ & $0.20$ & $0.84$ \\
$53.557$ & $88369$ & $5.222$ & $2.969$ & $5.685$ & $0.22$ & $1.09$ \\
$60.209$ & $90467$ & $5.216$ & $3.167$ & $5.782$ & $0.17$ & $1.15$ \\
$67.688$ & $92425$ & $5.211$ & $3.379$ & $5.875$ & $0.21$ & $1.08$ \\
$76.095$ & $94232$ & $5.206$ & $3.605$ & $5.965$ & $0.22$ & $0.93$ \\
$85.547$ & $95878$ & $5.200$ & $3.846$ & $6.052$ & $0.23$ & $0.69$ \\
$96.172$ & $97352$ & $5.195$ & $4.103$ & $6.134$ & $0.23$ & $0.64$ \\
$108.118$ & $98647$ & $5.189$ & $4.378$ & $6.214$ & $0.23$ & $0.65$ \\
$121.547$ & $99754$ & $5.184$ & $4.671$ & $6.289$ & $0.26$ & $0.53$ \\
$136.645$ & $100666$ & $5.179$ & $4.983$ & $6.361$ & $0.31$ & $0.45$ \\
$153.617$ & $101379$ & $5.173$ & $5.317$ & $6.430$ & $0.37$ & $0.51$ \\
$172.698$ & $102000$ & $5.168$ & $5.672$ & $6.497$ & $0.44$ & $0.59$ \\
$194.149$ & $102624$ & $5.162$ & $6.052$ & $6.563$ & $0.47$ & $0.60$ \\
$218.264$ & $103253$ & $5.157$ & $6.456$ & $6.630$ & $0.40$ & $0.49$ \\
$245.375$ & $103885$ & $5.152$ & $6.888$ & $6.697$ & $0.46$ & $0.54$ \\
$275.853$ & $104521$ & $5.146$ & $7.349$ & $6.764$ & $0.32$ & $0.38$ \\
$310.117$ & $105160$ & $5.141$ & $7.840$ & $6.831$ & $0.19$ & $0.26$ \\
$348.637$ & $105804$ & $5.136$ & $8.365$ & $6.898$ & $0.13$ & $0.26$ \\
$391.941$ & $106452$ & $5.130$ & $8.924$ & $6.964$ & $0.15$ & $0.39$ \\
$440.624$ & $107103$ & $5.125$ & $9.521$ & $7.031$ & $0.18$ & $0.43$ \\
$495.354$ & $107759$ & $5.119$ & $10.158$ & $7.098$ & $0.19$ & $0.41$ \\
$556.881$ & $108419$ & $5.114$ & $10.838$ & $7.165$ & $0.22$ & $0.44$ \\
$626.052$ & $109082$ & $5.109$ & $11.562$ & $7.232$ & $0.25$ & $0.48$ \\
$703.814$ & $109750$ & $5.103$ & $12.336$ & $7.299$ & $0.30$ & $6.53$ \\
$791.234$ & $110422$ & $5.098$ & $13.161$ & $7.365$ & $0.37$ & $7.67$ \\
$820.200$ & $110629$ & $5.096$ & $13.425$ & $7.386$ & $0.60$ & $4.59$ \\
\enddata
\end{deluxetable}


\bibliography{bibliography}{}
\bibliographystyle{aasjournal}

\end{document}